# Justice-informed Planning of Intermodal Autonomous Mobility-on-Demand Systems under Operational Constraints

G. Ganassoli, F. Mazzeo, C. Pasquale, S. Siri, M. Salazar

***Abstract*— To date, most of the research on transport planning has focused on optimizing revenues or utilitarian metrics such as average travel times, which often ends up penalizing the worst-off for the sake of profit or efficiency. At the same time, most of the research in transport justice has focused on assessing injustices, without being able to prescribe operational solutions. This paper contributes to bridging this gap and presents optimization models for justice-informed operational planning of intermodal mobility systems that explicitly account for the budget and safety limitations of users, and for infrastructural capacity constraints. Specifically, we first focus on an intermodal Autonomous Mobility-on-Demand (AMoD) system—where self-driving robotaxis provide on-demand mobility jointly with public transit and active modes—and characterize its operations from a mesoscopic planning perspective via network flow models. Second, we leverage these models to optimize system operations through both utilitarian efficiency and justice-informed objectives. We showcase our framework in a real-world case-study for Manhattan, New York. Our results show that monetary budgets significantly limit the social justice potential of AMoD systems if they are to be deployed as transportation network companies. At the same time, granting free public transit can result in sufficiency levels very close to a completely free intermodal AMoD system, where justice-informed operations can be achieved without compromising standard efficiency metrics, ultimately highlighting the strong potential of social policies.**

## I. Introduction

Urban mobility systems are experiencing a profound transformation driven by advances in autonomy and connectivity, enabling new coordinated transport services. Among these, Autonomous Mobility-on-Demand (AMoD) systems represent a paradigmatic shift, as fleets of self-driving vehicles provide on-demand mobility through centralized control. Whilst these technologies may appear beneficial, their large-scale deployment also raises questions about how mobility resources are allocated and who ultimately benefits from such improvements. When optimized using conventional metrics such as average travel time or operating cost, AMoD systems may reproduce or intensify existing inequalities, as already observed for other technology-driven mobility services [1]. Previous works have shown that combining AMoD with public transit and active modes, such as walking and cycling, can significantly improve system performance [2], even in the presence of self-interested users [3]. Nevertheless, most of the existing studies still adopt efficiency-oriented objectives that fail to capture the primary purpose of transport systems, namely, to provide accessibility [4].

G. Ganassoli, F. Mazzeo, C. Pasquale, S. Siri are with the Department of Informatics, Bioengineering, Robotics and Systems Engineering, University of Genoa, Genoa, Italy (e-mail: giacomo.ganassoli@edu.unige.it, cecilia.pasquale@unige.it, silvia.siri@unige.it). M. Salazar is with the Control Systems Technology section, Eindhoven University of Technology, Eindhoven, The Netherlands (e-mail: m.r.u.salazar@tue.nl).

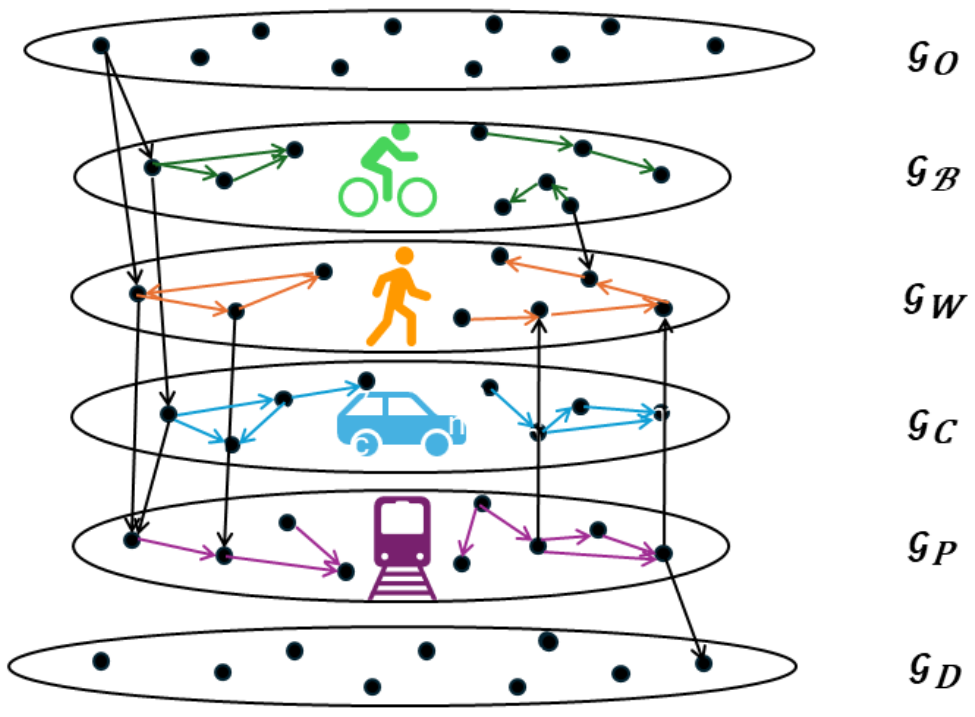

Fig. 1: Schematic of a multi-layer I-AMoD network.

Accessibility refers to the ability to perform essential activities in a reasonable manner [5], i.e., within reasonable time, effort, and cost. Incorporating accessibility into operational optimization is challenging under realistic conditions. Existing approaches typically assume unconstrained public transport capacity, treat congestion as exogenous, and neglect heterogeneous user conditions.

Against this backdrop, this paper proposes an optimization framework for the operational planning of Intermodal AMoD (I-AMoD) systems as shown in Fig. 1. Besides minimizing average travel times, in line with *Utilitarian Efficiency* paradigms, we also consider a sufficientarian approach aimed at achieving *Commute Sufficiency*, where travel times should not exceed a reasonable sufficiency threshold.

*Related Literature:* Our work relates to two main research streams: AMoD optimization models and transport justice. Several approaches have been proposed to study AMoD systems [6], [7]. In particular, multi-commodity network flow models [8] are well suited for planning and design, as they scale to large systems and can accommodate multiple objectives and constraints. They were applied to congestion-aware operations in [9], [10] and to ride-pooling in [11], [12]. Their interaction with the power grid was investigated in [13], while its application to intermodal systems was addressed in [2], [3], [14]. Yet all these contributions optimize time and/or costs in an aggregate utilitarian fashion that does not explicitly account for social justice.

The second stream of research focuses on principles of justice in transport systems [15]. In this context, accessibility-based metrics have been introduced to assess equity in transport systems [16], highlighting the limitations of utilitarian approaches [17], which tend to prioritize users with higher values of time and income [18]. To address their limitations, sufficientarianism was derived as an adequate distributive principle of justice for transportation in [4], requiring that all users reach a sufficient level of accessibility, while additional accessibility should be self-financing. This principle is measured through an accessibility sufficiency index further defined via the capabilities approach in [19]. However, all these approaches remain evaluative and are not suitable for actively prescribing planning solutions.

Perhaps closer to our work, cycling network upgrades are optimized in [20] following sufficientarian principles of justice, whilst different principles of justice are operationalized for bus frequency optimization in [21], [22]. However, none of these papers considers more general intermodal settings. A recent study [23] presented a first step in leveraging optimization models to study different formulations of transport justice for operational planning. However, this initial work is built on a number of assumptions that do not generalize its applicability, such as negligible endogenous congestion effects, sufficient cycling safety, and a population that can afford access to all modes of transportation.

In conclusion, to the best of the authors' knowledge, there are no frameworks to optimize the operational planning of intermodal mobility systems in a justice-informed manner and under monetary, safety, and infrastructural constraints.

*Statement of Contributions:* This paper presents optimization models to plan the operation of I-AMoD systems in line with justice-informed objectives accounting for user-specific and infrastructural constraints. First, we propose a modeling framework accounting for user-related limitations in monetary mobility budgets and cycling safety, as well as road and public transit capacity limits. Second, we introduce and formalize two distinct operational paradigms: one based in utilitarian efficiency objectives (i.e., average travel time minimization) and one grounded in a justice-informed metric capturing commute sufficiency. Finally, we validate the proposed framework through a real-world case study in Manhattan, New York City.

*Organization:* Section II introduces the extended network flow model for I-AMoD systems, based on which utilitarian efficiency and commute sufficiency problems are formulated. Section III presents the validation of our approach through a case study in the Manhattan borough of New York City. Finally, Section IV concludes the paper, summarizing the key results of the case study.

## II. Methodology

This section presents a network flow model of the I-AMoD system and leverages it to formulate optimization problems aimed at optimizing the operational planning of the system in line with different operational paradigms.

### A. Intermodal Network Structure

In line with [23], we represent the mobility system and its various transportation modes using a directed graph $\mathcal{G} = (\mathcal{V}, \mathcal{A})$, where $\mathcal{V}$ is the set of nodes and $\mathcal{A} \subseteq \mathcal{V} \times \mathcal{V}$ is the set of arcs. The graph incorporates several layers: a bicycle layer $\mathcal{G}_\mathrm{B} = (\mathcal{V}_\mathrm{B}, \mathcal{A}_\mathrm{B})$, a walking layer $\mathcal{G}_\mathrm{W} = (\mathcal{V}_\mathrm{W}, \mathcal{A}_\mathrm{W})$, an AMoD network layer $\mathcal{G}_\mathrm{C} = (\mathcal{V}_\mathrm{C}, \mathcal{A}_\mathrm{C})$, a public transport layer $\mathcal{G}_\mathrm{P} = (\mathcal{V}_\mathrm{P}, \mathcal{A}_\mathrm{P})$, and two additional layers, the origin layer $\mathcal{G}_\mathrm{O} = (\mathcal{V}_\mathrm{O}, \emptyset)$ and the destination layer $\mathcal{G}_\mathrm{D} = (\mathcal{V}_\mathrm{D}, \emptyset)$, which contain all origin and destination nodes, respectively, and no internal connections. In the considered graph, no nodes are shared between the different layers. More specifically, the walking and bicycle networks are modeled through pedestrian streets and bike lanes $(i,j) \in \mathcal{A}_\mathrm{W}$ and $(i,j) \in \mathcal{A}_\mathrm{B}$, connecting intersections $i \in \mathcal{V}_\mathrm{W}$ and $i \in \mathcal{V}_\mathrm{B}$, respectively. The AMoD network consists of nodes $i \in \mathcal{V}_\mathrm{C}$ and road segments $(i,j) \in \mathcal{A}_\mathrm{C}$. The public transport system, which may include trams and subway lines, is modeled by station nodes $i \in \mathcal{V}_\mathrm{P}$ and line segments $(i,j) \in \mathcal{A}_\mathrm{P}$. This multilayer graph is illustrated in Figure 1.

To model the possibility of customers switching between transportation modes (e.g., exiting a subway and taking an AMoD ride) we define a set of mode-switching arcs $\mathcal{A}_\mathrm{S}$:

$$\begin{aligned}\mathcal{A}_\mathrm{S} =& (\mathcal{V}_\mathrm{O} \times (\mathcal{V}_\mathrm{B} \cup \mathcal{V}_\mathrm{W} \cup \mathcal{V}_\mathrm{C} \cup \mathcal{V}_\mathrm{P})) \\ & \cup (\mathcal{V}_\mathrm{B} \times (\mathcal{V}_\mathrm{W} \cup \mathcal{V}_\mathrm{C} \cup \mathcal{V}_\mathrm{P} \cup \mathcal{V}_\mathrm{D})) \\ & \cup (\mathcal{V}_\mathrm{W} \times (\mathcal{V}_\mathrm{B} \cup \mathcal{V}_\mathrm{C} \cup \mathcal{V}_\mathrm{P} \cup \mathcal{V}_\mathrm{D})) \\ & \cup (\mathcal{V}_\mathrm{C} \times (\mathcal{V}_\mathrm{B} \cup \mathcal{V}_\mathrm{W} \cup \mathcal{V}_\mathrm{P} \cup \mathcal{V}_\mathrm{D})) \\ & \cup (\mathcal{V}_\mathrm{P} \times (\mathcal{V}_\mathrm{B} \cup \mathcal{V}_\mathrm{W} \cup \mathcal{V}_\mathrm{C} \cup \mathcal{V}_\mathrm{D})).\end{aligned}$$

Furthermore, we consider a set of $R$ regions $r \in \mathcal{R} = \{1, \ldots, R\}$ with population $n_\mathrm{r}$ such that $\sum_{r \in R} n_\mathrm{r} = n_\mathrm{pop}$, where $n_\mathrm{pop}$ denotes the total population. Each region is associated with a set of mobility demands $\mathcal{M}_\mathrm{r}$ referring to trips from locations in region $r$ to other locations (e.g., from a home to a workplace). These are formally represented by the tuple $(o_\mathrm{m}, d_\mathrm{m}, \alpha_\mathrm{m})$ with $m \in \mathcal{M}_\mathrm{r}$, where $o_\mathrm{m} \in \mathcal{V}_\mathrm{O}$ is the origin node belonging to region $r$, $d_\mathrm{m} \in \mathcal{V}_\mathrm{D}$ is the destination node and $\alpha_\mathrm{m}$ denotes the number of users per unit of time. Moreover, for each region $r$, the demand set $\mathcal{M}_\mathrm{r}$ is subdivided into two disjoint subsets $\mathcal{M}_\mathrm{r}^1$ and $\mathcal{M}_\mathrm{r}^2$, such that $\mathcal{M}_\mathrm{r}^1 \cup \mathcal{M}_\mathrm{r}^2 = \mathcal{M}_\mathrm{r}$. In particular, $\mathcal{M}_\mathrm{r}^1$ refers to the mobility demand that can be satisfied through the use of bicycles, while $\mathcal{M}_\mathrm{r}^2$ refers to the demand without the use of bicycles to account for people who are not capable of cycling.

Finally, a constant travel time vector $t_\mathrm{a} \geq 0$ is introduced to define the time to traverse each arc $a \in \mathcal{A}$, representing both the movement with a mode (AMoD, bike, public transport, or walking) and the switching arcs (e.g., parking a bike, exiting an AMoD ride, waiting, or boarding a public transport vehicle).

### B. Operational Constraints

Building upon the network defined so far, we introduce arc-specific flow variables for each origin-destination pair, denoted as $x_a^m \geq 0$, which quantify the number of individuals associated with demand $m \in \mathcal{M}_r$ of region $r \in \mathcal{R}$ traveling along arc $a \in \mathcal{A}$ per unit time. Furthermore, we define the rebalancing flows of empty vehicles as $x_a^\mathrm{R} \geq 0$ for $a \in \mathcal{A}_\mathrm{C}$.

In order to ensure that flow is not assigned to cycling segments that could pose risks to users, we eliminate all arcs in $\mathcal{A}_\mathrm{B}$ with safety levels below the established threshold:

$$s_a \leq S \quad \forall a \in \mathcal{A}_\mathrm{B}, \tag{1}$$

where $s_a$ represents the unsafety level of a specific cycling arc $a$ and $S$ is the maximum acceptable unsafety threshold.

Flow conservation in the network must be ensured by imposing that the flow entering node $j$ is equal to the flow exiting the same node. This must be guaranteed both for mobility demands that can be satisfied using any transportation mode and for those that do not allow the use of bikes:

$$\alpha_m 1_{j=o_m} + \sum_{i:(i,j)\in\mathcal{A}} x^m_{(i,j)} = \alpha_m 1_{j=d_m} + \sum_{k:(j,k)\in\mathcal{A}} x^m_{(j,k)} \quad \forall j \in \mathcal{V}, m \in \mathcal{M}^1_r, r \in \mathcal{R} \tag{2}$$

$$\alpha_m 1_{j=o_m} + \sum_{i:(i,j)\in\mathcal{A}\setminus\mathcal{A}_\mathrm{B}} x^m_{(i,j)} = \alpha_m 1_{j=d_m} + \sum_{k:(j,k)\in\mathcal{A}\setminus\mathcal{A}_\mathrm{B}} x^m_{(j,k)} \quad j \in \mathcal{V}\setminus\mathcal{V}_\mathrm{B}, m \in \mathcal{M}^2_r, r \in \mathcal{R}, \tag{3}$$

where $1$ is a Boolean indicator function defined as $1_b = 1$ if $b$ is true and 0 otherwise.

We ensure the flow balance of AMoD vehicles, guaranteeing that the total AMoD flow (both empty and occupied) entering any node in the set $\mathcal{V}_\mathrm{c}$ equals the flow exiting the same node:

$$\sum_{i:(i,j)\in\mathcal{A}_\mathrm{c}} \left( x^\mathrm{R}_{(i,j)} + \sum_{r\in\mathcal{R}} \sum_{m\in\mathcal{M}_r} x^m_{(i,j)} \right) = \sum_{k:(j,k)\in\mathcal{A}_\mathrm{c}} \left( x^\mathrm{R}_{(j,k)} + \sum_{r\in\mathcal{R}} \sum_{m\in\mathcal{M}_r} x^m_{(j,k)} \right) \quad \forall j \in \mathcal{V}_\mathrm{c}. \tag{4}$$

We impose that the total number of AMoD vehicles, both occupied and empty, on the road arcs remains below a certain threshold:

$$\sum_{a\in\mathcal{A}_\mathrm{c}} t_a \cdot \left( x^\mathrm{R}_a + \sum_{r\in\mathcal{R}} \sum_{m\in\mathcal{M}_r} x^m_a \right) \leq N^\mathrm{max}_\mathrm{AMoD}, \tag{5}$$

where $N^\mathrm{max}_\mathrm{AMoD}$ denotes the maximum number of AMoD vehicles allocable within the network.

We ensure that, for every region and every mobility demand, the average transportation cost is below the average daily personal budget:

$$\frac{\sum\limits_{a\in\mathcal{A}} (c_a \cdot x^m_a)}{\alpha_m} \leq B_r \quad \forall m \in \mathcal{M}_r, r \in \mathcal{R}, \tag{6}$$

where $B_r$ represents the average daily personal budget for region $r$ and $c_a$ is the unit cost of crossing arc $a$.

An upper bound on the flow of AMoD vehicles allocated on each road arc $a \in \mathcal{A}_\mathrm{c}$ is considered to limit the endogenous congestion effects generated by the AMoD service:

$$x^\mathrm{R}_a + \sum_{r\in\mathcal{R}} \sum_{m\in\mathcal{M}_r} x^m_a \leq f^\mathrm{max}_a \quad \forall a \in \mathcal{A}_\mathrm{c}, \tag{7}$$

where $f^\mathrm{max}_a$ represents the maximum admissible flow of AMoD vehicles on arc $a \in \mathcal{A}_\mathrm{c}$. This constraint allows to preserve the assumption of constant travel times $t_a$ by preventing excessive traffic levels caused by the AMoD fleet.

Similarly, we ensure that for each arc $a$ in the public transport network, the total number of users per unit of time does not exceed the capacity of that arc:

$$\sum_{r\in\mathcal{R}} \sum_{m\in\mathcal{M}_r} x^m_a \leq K_a \quad \forall a \in \mathcal{A}_\mathrm{p}, \tag{8}$$

where $K_a$ represents the capacity of arc $a$.

### C. Utilitarian Efficiency Optimization Problem

The Utilitarian Efficiency problem can be defined through the following objective function, which minimizes the total travel time in the system:

$$J_\mathrm{Util,Eff} = \sum_{a\in\mathcal{A}} t_a \cdot \sum_{r\in\mathcal{R}} \sum_{m\in\mathcal{M}_r} x^m_a + \gamma^\mathrm{R} \cdot \sum_{a\in\mathcal{A}_\mathrm{C}} t_a \cdot x^\mathrm{R}_a, \tag{9}$$

where $\gamma^\mathrm{R}$ is a small regularization term for rebalancing.

**Problem 1** (Utilitarian Efficiency)**.** *The optimal flows minimizing the average travel time of the population are obtained by solving the following optimization problem:*

$$\min_{\{x^m\}_m \in \mathbb{R}^{|\mathcal{A}|}, x^\mathrm{R} \in \mathbb{R}^{|\mathcal{A}_\mathrm{C}|}} J_\mathrm{Util,Eff} \quad \text{s.t. (2) - (8).}$$

Problem 1 can be efficiently solved as a Linear Program (LP). However, it is limited in scope, focusing solely on minimizing the average travel time for all users, without considering transportation justice aspects, such as accessibility fairness.

### D. Commute Sufficiency Optimization Problem

Accessibility is a multifaceted concept related to accommodating travel demands in a reasonable way [4]. To quantify, a reasonable travel time threshold $T_\mathrm{suff}$ is considered adopting a sufficientarian perspective, and the inaccessibility of an O-D pair is measured by the additional time required to reach the destination beyond this threshold, using a slack variable $\varepsilon_m$:

$$\varepsilon_m = \max\left\{0, \frac{\sum_{a\in\mathcal{A}}(t_a \cdot x^m_a)}{\alpha_m} - T_\mathrm{suff}\right\} \quad \forall m \in \mathcal{M}_r, r \in \mathcal{R}, \tag{10}$$

which can be relaxed to convex as

$$\varepsilon_m \geq 0, \quad \forall m \in \mathcal{M}_r, r \in \mathcal{R}, \tag{11}$$

$$\varepsilon_m \geq \frac{\sum_{a\in\mathcal{A}}(t_a \cdot x^m_a)}{\alpha_m} - T_\mathrm{suff}, \quad \forall m \in \mathcal{M}_r, r \in \mathcal{R}. \tag{12}$$

This relaxation is lossless, as $\varepsilon_\mathrm{m}$ will be subject to minimization, as shown below.

The commute insufficiency metric was developed drawing inspiration from the $\mathrm{FGT}_2$ poverty index [16], as it accounts for both the extent and the severity of a deficiency. Importantly, such a formulation penalizes excessive travel times for each O-D pair *on average*, since trip $m$ could be realized along different paths with different individual travel times. Such an operational model could be realized for repetitive daily commutes with turn-taking mechanisms, e.g., via token economies [24], in line with equity over time paradigms [21].Consequently, we define $u_\mathrm{r}$ as the commute insufficiency level for region $r$ as

$$u_r = \frac{\sum_{m\in\mathcal{M}_r} \alpha_m \cdot \varepsilon_m^2}{\sum_{m\in\mathcal{M}_r} \alpha_m} \quad \forall r \in \mathcal{R}, \tag{13}$$

where $\varepsilon_\mathrm{m}$ is squared to give greater influence to larger deviations from $T_\mathrm{suff}$ in the regional commute insufficiency level. Now we can define the objective function to capture the commute insufficiency experienced by the entire population as

$$J_\mathrm{Comm,Suff} = \frac{\sum_{r\in\mathcal{R}} n_r \cdot u_r}{\sum_{r\in\mathcal{R}} n_r}. \tag{14}$$

**Problem 2** (Commute Sufficiency)**.** *The optimal flows minimizing the commute insufficiency of the population result from*

$$\min_{\{x^m\}_m\in\mathbb{R}^{|\mathcal{A}|}, x^\mathrm{R}\in\mathbb{R}^{|\mathcal{A}_\mathrm{C}|}} J_\mathrm{Comm,Suff} + \gamma^\mathrm{time} \cdot J_\mathrm{Util,Eff}$$
$$\text{s.t. (2) - (8), (11) - (13),}$$

*where $\gamma^\mathrm{time}$ is a small regularization term to avoid self-loops that can happen for accessible O-D pairs.*

Problem 2 is a Quadratic Program (QP) that can be solved with standard solvers with global optimality guarantees.

Notably, once the optimal flows are obtained from Problems 1 and 2, it is possible to reconstruct explicit paths for each O-D pair through a dedicated path allocation procedure. Such an algorithm is omitted here due space limitations and can be found in [25].

## III. Results of Manhattan Case Study

This section presents a real-world case-study for Manhattan, NYC. The multimodal network was obtained from OpenStreetMap, following the approach in [3]. Road and bicycle layers each comprise 1,351 nodes and 3,137 arcs, while the walking layer uses the same node set with 4,330 arcs to capture pedestrian pathways. The public transport layer models the Manhattan subway system with 121 nodes and 334 arcs. Origin and destination layers were created from the same node set to represent trip endpoints.

We built the demand matrix from the Longitudinal Employer-Household Dynamics (LEHD) program of the United States Census Bureau of Statistics [26]. From the full set of NYC home-to-work commutes, only trips with both origin and destination in Manhattan were retained. Data were aggregated by district, spatially matched to the nearest network node, and then re-aggregated into the 39 administrative regions of Manhattan, yielding 1,511 origin–destination pairs; 22.7% of commuters were removed to account for car ownership [27]. Moreover, to satisfy the transportation alternatives constraints, the demand matrix was split to account for those who may have difficulty using bicycles. The splitting share was derived from the regional proportion of residents above a certain age, which is introduced as a modeling proxy for potential difficulties in using bicycle-based services [28]. Finally, the transportation costs for both AMoD and the subway were estimated using the actual fares of public transit and taxi services in Manhattan [29], [30].

The main parameters adopted for the Manhattan case study are summarized in Table I. All optimization problems were parsed with YALMIP [31] and solved using Gurobi. On the TU/e Umbrella HPC Cluster (using 32 CPU cores and 200 GB RAM), the solution of each optimization problem required approximately 55 min. The following sections compare these outcomes in terms of travel time and accessibility, focusing on the impact of the budget constraints and the introduction of free public transport.

### A. Utilitarian Efficiency vs. Commute Sufficiency

The results of the two proposed operational models are reported in Table II and Fig. 2, where the latter shows histograms of the commute travel time distribution.

Fig. 2a shows the performance of the standard Utilitarian Efficiency model, which focuses on minimizing travel times, resulting in an average travel time of 14.26 min, and a commute insufficiency of 8.51 min$^2$. Fig. 2b shows the performance of the Commute Sufficiency approach, which instead aims at reducing travel times exceeding the $T_\mathrm{suff}$ threshold. This operation results in a commute insufficiency level of 6.69 min$^2$, more than 20% lower compared to the Utilitarian Efficiency model, whilst only slightly increasing the average travel time to 14.63 min, i.e., less than 3%. Fig. 3 shows heatmaps capturing insufficiency levels from a geographical perspective, revealing how they are mostly present at the outskirts of the peninsula and how planning the operations for commute sufficiency can only partly address such issues. As shown in Section III-B below, the reason for the presence of injustice in operations is to be found mainly in limited monetary budgets of the population w.r.t. the cost of the different modes.

### B. Impact of Budget Constraints

In this section, we analyze the impact of the budget constraints on the model, with specific reference to the Commute Sufficiency formulation, by assuming a completely free intermodal AMoD system. Fig. 2c shows that the share of trips served by AMoD and subway significantly increase when budget constraints are lifted, leading to a substantial improvement both in terms of commute sufficiency and travel times. In particular, the commute insufficiency level drastically decreases to $0.34\,\mathrm{min}^2$, whilst the average travel time is reduced by more than 10% to 12.35 min. When examining the commute insufficiency heatmap in Fig. 3c, the situation is significantly improved also in the most peripheral areas that are less connected and characterized by low income levels.

TABLE I: Parameters for the case study.

| Parameter | Value | Unit |
|---|---|---|
| $N_\mathrm{AMoD}^\mathrm{max}$ | 24,000 | vehicles |
| $\sum_m \alpha_\mathrm{m}$ | 445,851 | users/day |
| $T_\mathrm{suff}$ | 20 | min |

TABLE II: Optimization results for the two problems

| Objective | Util,Eff | Comm,Suff |
|---|---|---|
| Average Travel Time [$min$] | 14.26 | 14.63 |
| Commute Insufficiency [$min^2$] | 8.51 | 6.69 |

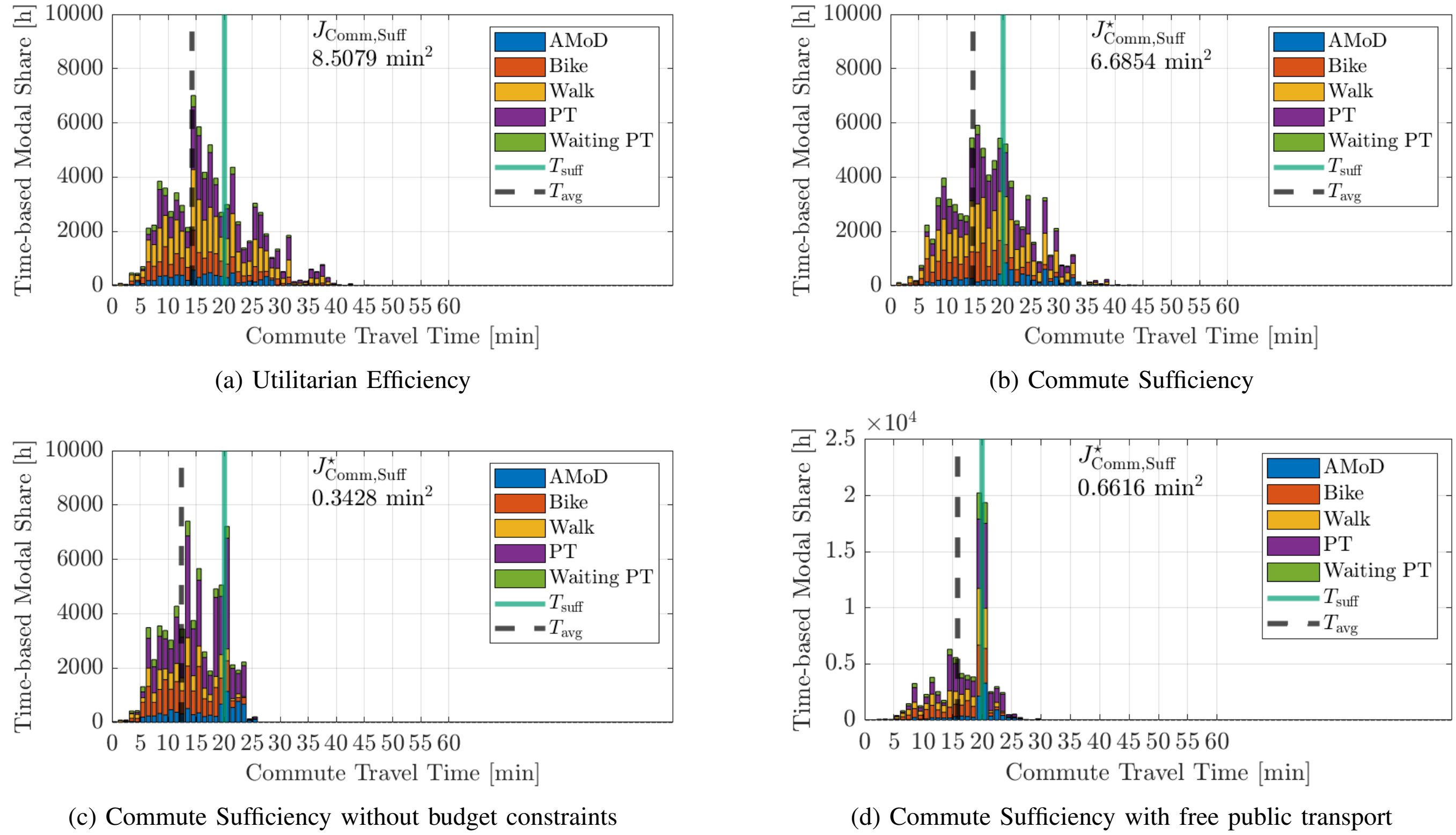

Fig. 2: Distribution of the commute travel time and mode in Manhattan when optimizing for Utilitarian Efficiency (nominal scenario) and Commute Sufficiency under three scenarios (nominal, without budget constraints and with free public transport).

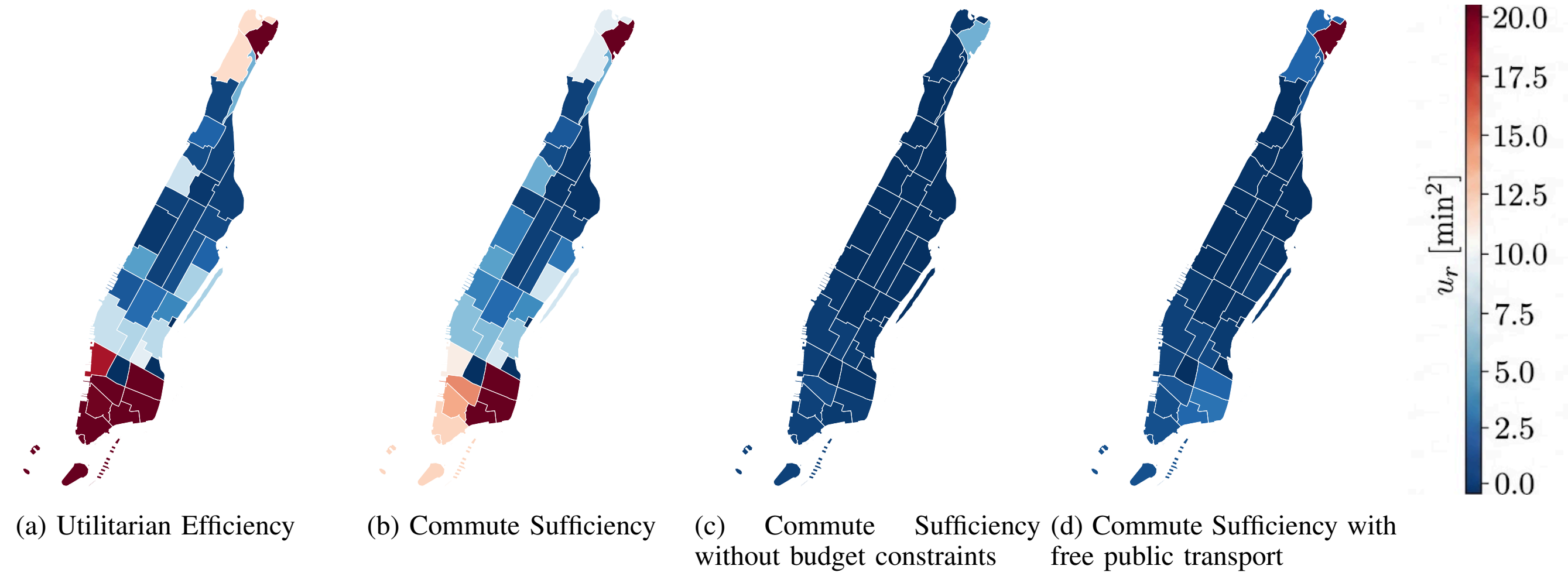

Fig. 3: Commute insufficiency levels in Manhattan when optimizing for Utilitarian Efficiency (nominal scenario) and Commute Sufficiency under three scenarios (nominal, without budget constraints and with free public transport).

### C. Impact of Free Public Transport

Finally, we analyze the impact of making public transport free, as recently proposed by New York City Mayor Zohran Mamdani [32]. Fig. 2d shows that providing free subway access clearly increases the use of public transport, resulting in more trips being completed within the maximum time threshold $T_{\mathrm{suff}}$, and consequently improving commute sufficiency. In particular, the commute insufficiency levels decrease to $0.66\,\mathrm{min}^2$, corresponding to a 90% reduction compared to the original scenario, whilst almost matching the levels obtained with a fully free I-AMoD system.

Such an improvement is also visible in Fig. 3d, where significant gains are again observed in the peripheral regions.

## IV. Conclusion

In this work, we presented optimization models to plan the operations of intermodal mobility systems in a justice-informed manner, whilst explicitly accounting for the impact of infrastructural and user-related constraints. We compared two operational models, one focused on utilitarian efficiency objectives of average travel time minimization, and a justice-informed one aiming at reducing exceedingly long travel times in a sufficientarian manner. Our results show that fairer outcomes come at a negligible cost to efficiency. Moreover, we showed that budget limitations significantly affect overall efficiency and justice levels, highlighting that benefits of justice-oriented operations are limited when AMoD systems are deployed via standard transportation network companies. Finally, we found that social policies such as free public transit provision can significantly improve justice levels for most users.

This work attempted another step outside of the "engineering trap" and should be extended in several dimensions. First, quantitative models rely on data that often suffer from structural biases, a source of injustice that must be addressed before designing more just transport systems [33]. Second, we would like to challenge the conventional technical view of transport as "a matter of simply moving people and things across space" [34]. To this end, we aim at accounting for broader aspects of human mobility, including different dimensions of justice, embodied mobility experiences, and community wellbeing at large.

## Acknowledgements

We thank Dr. I. New for proofreading this paper and for re-humanizing sections where LLMs had been used to improve grammar and flow.